\documentstyle[prl,aps,graphicx]{revtex}

\DeclareMathAlphabet{\bi}{OML}{cmm}{b}{it}

\begin{document}

\draft

\twocolumn[\hsize\textwidth\columnwidth\hsize\csname@twocolumnfalse\endcsname

\title{Charge degree of freedom and single-spin fluid model in
$\mathbf{YBa_{2}Cu_{4}O_{8}}$}
\author{A. Suter\cite{Auth1}, M. Mali, J. Roos, and D. Brinkmann}
\address{Physik-Institut, Universit\"{a}t Z\"{u}rich,
CH-8057 Z\"{u}rich, Switzerland}

\maketitle
\date{\today}

\begin{abstract}
We present a $^{17}$O nuclear magnetic resonance study in the
stoichiometric superconductor $\mathrm{YBa_{2}Cu_{4}O_{8}}$.  A
double irradiation method enables us to show that, below around
200~K, the spin-lattice relaxation rate of plane oxygen is not
only driven by magnetic, but also significantly by quadrupolar
fluctuations, i.e.\ low-frequency charge fluctuations.  In the
superconducting state, on lowering the temperature, the
quadrupolar relaxation diminishes faster than the magnetic one.
These findings show that, with the opening of the pseudo spin gap,
a charge degree of freedom of mainly oxygen character is present
in the electronic low-energy excitation spectrum.
\end{abstract}

]


One of the central issues in understanding cuprate-based
high-temperature superconductors (HTSC) has been to determine the
minimal number of electronic degrees of freedom which are necessary to
describe the physics at the atomic scale of these structures.  Zhang
and Rice proposed that doped holes, which go into oxygen states, form
a spin-resonant singlet state with the quasi-localized holes at copper
\cite{zhang88}.  Therefore, only {\em one} spin degree of freedom
would be necessary to describe the low-energy dynamics of the
electronic system in the normal state of the HTSC. This scenario is
often called {\it single-spin fluid} (SSF) and is described within a
$t$-$J$ model.  However, as was discussed by various authors
\cite{emery88lu90}, this model might be an oversimplification.

Due to its local character, nuclear magnetic resonance (NMR) is a very
attractive method to address these questions.  NMR probes the
low-energy excitations of the HTSC electronic system and it provides
information about copper and oxygen independently.  Studies of the
{\it uniform} spin susceptibility, Re$\{ \chi(q=0,\omega=0 \}$
\cite{takigawa91bankay94}, indicate that only one spin degree of
freedom is involved thus supporting the SSF model.  On the other hand,
oxygen and copper reveal a very different temperature dependence of
the spin-lattice relaxation \cite{hammel89mangelschots93} which is
related to the {\it dynamic} spin susceptibility,
Im$\{\chi(q,\omega_{L}) \}$ \cite{moriya63}.  This could suggest
independent spin degrees of freedom.  However, as pointed out before
\cite{mila89shastry89bulut90}, the hyperfine field, due to
antiferromagnetically correlated copper spins, cancels at the oxygen
site and this fact can explain the different temperature dependences
within a SSF model.

There is, however, evidence that this view is incomplete.  The
$^{17}$O NMR in Y-Ba-Cu-O HTSC has extensively been studied by various
groups.  Especially the pronounced temperature dependence of the
anisotropy of the experimentally determined effective spin-lattice
relaxation rate, $^{17}W_{\rm eff}$, of the plane oxygen is still
puzzling \cite{barriquand91Horvatic93,suter97,martindale98} while the
SSF model predicts an almost temperature independent rate anisotropy.
(The meaning of {\it effective} will be discussed below.)  As we
pointed out in Ref.\ \cite{suter97}, it is also very difficult to
explain, within a simple SSF model, the ratio of this rate and the
yttrium rate, namely $^{17}W_{\rm eff}/{^{89}W}$ (with the external
magnetic field $B_0$ parallel to the $c$ axis).  Furthermore, in a
detailed analysis, Walstedt {\em et al.} \cite{walstedt94walstedt95}
compared their $^{17}$O and $^{63}$Cu NMR results from
$\mathrm{La_{2-x}Sr_xCuO_4}$ studies with inelastic neutron scattering
data and concluded that a single-band picture is inadequate.

In this Letter, we will show that the plane oxygen spin-lattice
relaxation in $\mathrm{YBa_{2}Cu_{4}O_{8}}$, below about 200~K, is not
only driven by magnetic but also by quadrupolar fluctuations, $i.e.$
low-frequency charge fluctuations.  This fact implies that in the
temperature region where the pseudo spin gap is present, a {\em
charge} degree of freedom has to be taken into account, in addition to
the spin degree of freedom; hence the simple SSF model is inadequate.

Our results were obtained by a double irradiation technique which we
developed to extract the quadrupolar contribution to the overall NMR
relaxation \cite{suter99}; we sketch this technique as follows.  The
relaxation of the spin system towards its thermodynamic equilibrium is
described by the so-called master equation \cite{abragam61} and, in
the interaction representation \cite{andrew61,suter98}, it is given by
\begin{equation} \label{eq:master-eq}
  \frac{d}{dt} {\mathbf{P}}(t) = {\mathbf{R}}(W, W_{1}, W_{2}) [
{\mathbf{P}}(t) -
         {\mathbf{P}}(0) ] + {\mathbf{S}}_{\rm rf} {\mathbf{P}}(t).
\end{equation}

\begin{figure}[h]
        \centering
        \includegraphics[width=0.8\linewidth]{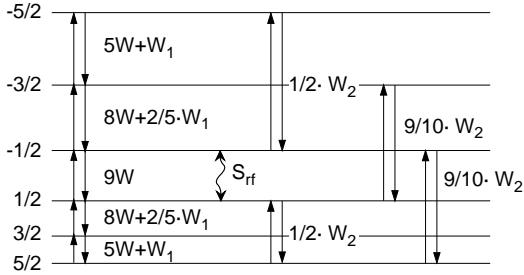}
        \caption{Probabilities for transitions between the spin
        energy levels effected by magnetic ($W$) and quadrupolar
        ($W_{1},W_{2}$) spin-lattice relaxation processes for
         $I=5/2$. $S_{\rm rf}$ denotes transitions due to a
         stimulating radio-frequency field.}
        \label{fig:level-diag}
\end{figure}

\noindent Here, ${\mathbf{P}}(t)$ is the nuclear spin system
population vector of the different energy levels with
${\mathbf{P}}(0)$ being the equilibrium value.  The relaxation matrix
${\mathbf{R}}(W, W_{1}, W_{2})$ is, in general, a function of the
magnetic relaxation rate $W$ (causing transitions with $| \Delta m |=
1$) and the two quadrupolar relaxation rates $W_{1}$ ($| \Delta m | =
1$) and $W_{2}$ ($| \Delta m | = 2$) \cite{andrew61,suter98}.  The
transition probabilities for a spin $I=5/2$ (like $^{17}$O) are
sketched in Fig.\ \ref{fig:level-diag}.  ${\mathbf{S}}_{\rm rf}$
denotes transitions due to an additional stimulating radio-frequency
(rf) field which is essential for the separation of magnetic from
quadrupolar contributions to the relaxation, as will be discussed
below.  In HTSC, the Zeeman degeneracy is lifted due to the crystal
field so that different transitions have distinguishable resonance
frequencies \cite{abragam61}.  In case the relaxation is purely
magnetic it is easy to show that the saturation of any line does not
affect the intensities of others.  This is no more true for the
general case where $W$ as well as $W_{1}$ and $W_{2}$ contribute to
the relaxation.

In the limit of dynamical saturation (i.e.\ $d{\mathbf{P}}(t)/dt =
0$), Eq.\ (\ref{eq:master-eq}) can be solved exactly \cite{suter99}.
The essential result is that, mainly due to the $| \Delta m | = 2$
transitions, the intensities of different lines change compared to a
standard detection method ($e.g.$ spin-echo sequence) where $S_{\rm
rf}$ is zero.  In our present experiment, $S_{\rm rf}$ saturates the
$(-1/2, 1/2)$ transition and we observe the other transitions,
yielding signals with intensities $I_{\rm dyn}$.  The spin-echo
sequence (with $S_{\rm rf}$=0) provides signals with intensities
$I_{\rm se}$.  For the $(-3/2,-1/2)$ transition the {\em enhancement}
factor, $E$, is then given by \cite{suter99}:

\begin{equation}
    E = \frac{I_{\rm dyn}}{I_{\rm se}} = 1 + \frac{\mu}{\zeta},
\end{equation}
with $\mu = 45W_{2} (10W + 2W_{1} + W_{2})$ and
     $\zeta = 4000W^2 \\
      + 1000WW_{1} + 40 W_{1}^2
      + 1100WW_{2} + 160W_{1}W_{2} + 45W_{2}^2$.

\noindent $E$ is a rather insensitive function with respect to
$W_{1}$, since $W_{1}$ connects the same energy levels as $W$, except
the $(-1/2, 1/2)$ transition \cite{suter99}.  However, $E$ depends
significantly on $W_{2}$ which makes the separate detection of
quadrupole relaxation possible.  The enhancement $E$ for various
transitions is quite different; for the $(-5/2, -3/2)$ transition,
there is almost no enhancement.  More information and details
concerning the method and its experimental realization as well as
different cross-checks of the results are given in Ref.\
\cite{suter99}.

\begin{figure}[h]
        \centering
        \includegraphics[width=\linewidth]{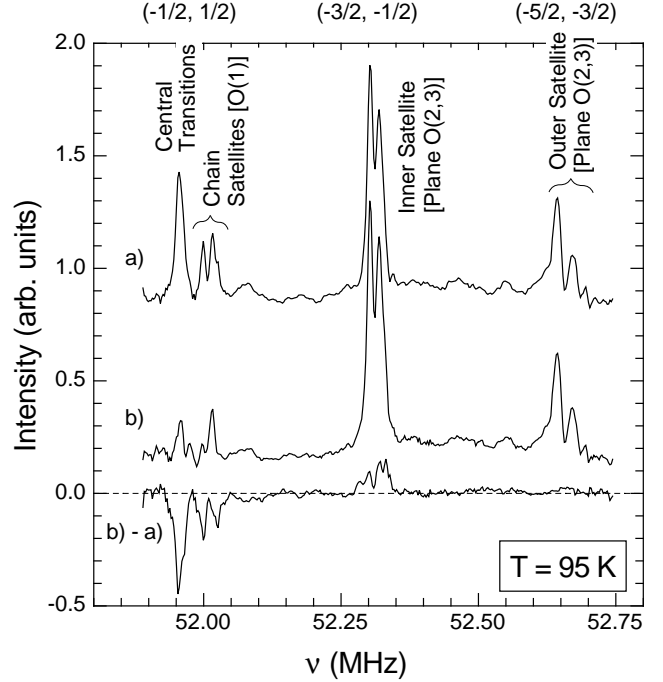}
        \caption{(a) $^{17}$O central transition and high frequency
                 satellites of chain and plane oxygen in
                 $\mathrm{YBa_{2}Cu_{4}O_{8}}$, respectively,
                 obtained by a standard spin-echo experiment.
                 (b) $^{17}$O spectrum obtained after dynamic
                 saturation of the ($-1/2,1/2$) transition.
                 Bottom: Difference of spectrum (b) and (a).}
        \label{fig:diffspec}
\end{figure}

The measurements were performed on an oriented and $^{17}$O enriched
$\mathrm{YBa_{2}Cu_{4}O_{8}}$ powder sample, used in previous studies
\cite{suter97}, in a field of $8.9945~{\mathrm{T}}$ applied along the
c axis; Fig.\ \ref{fig:diffspec} presents the results for $T=95$~K.
The chain sites O(1) are not of interest in this work and therefore
will not be discussed.  Also, we are not concerned with the plane
oxygen satellites O(2,3) splitting which is due to the orthorhombic
symmetry \cite{mangelschots92a} of the crystal.  Striking the eye is
the negative intensity of the central transition $(-1/2, 1/2)$ in the
difference spectrum, which means saturation of this transition.  This
is in contrast with the central and the satellite transitions of {\it
chain} oxygen which are not saturated because O(1) relax considerably
faster than plane nuclei for which the pulse sequence was optimized.
That this explanation is correct has been proven by the symmetric
experiment where we dynamically saturated the outer high-frequency
satellite \cite{suter99}.

The most important feature of the difference spectrum is the remaining
positive intensity at the position of the $(-3/2,-1/2)$ transition.
This intensity enhancement clearly shows that there is a quadrupolar
contribution to the relaxation of plane oxygen nuclei.  At the
position of the $(-5/2, -3/2)$ transition, the intensity is almost
zero, as expected from our calculations.

\begin{figure}[h]
       \centering
        \includegraphics[width=\linewidth]{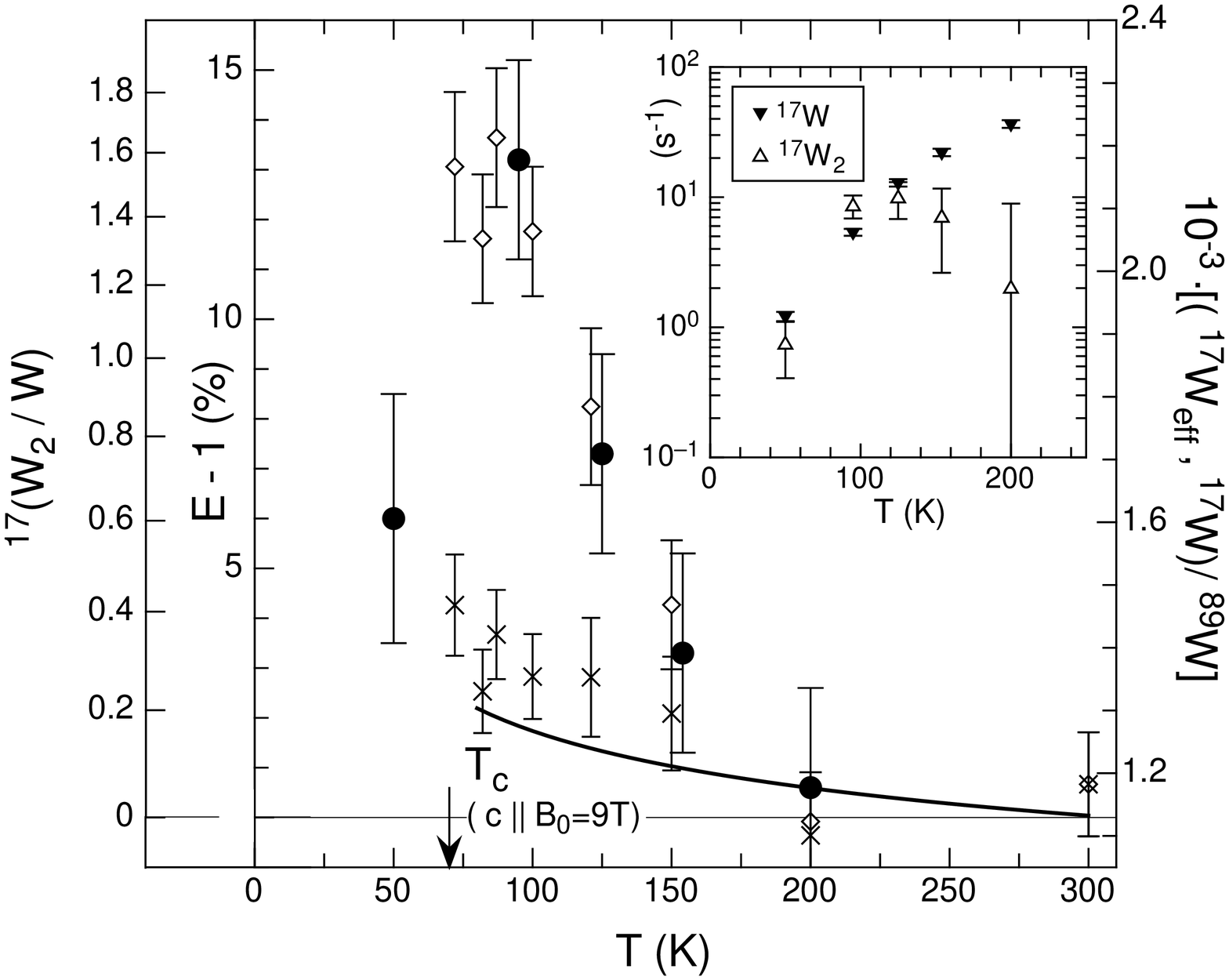}
        \caption
         {Temperature dependence of intensity enhancement of the
         $^{17}$O $(-3/2, -1/2)$ transition ($\bullet$);
         ${^{17}W}_{\rm eff} / {^{89}W}$ ($\diamond$);
         ${^{17}W} / {^{89}W}$ ($\times$); theoretical prediction
         for ${^{17}W} / {^{89}W}$ (solid curve)
         \protect\cite{millis90zha96}. Inset:
         $^{17}W$ and $^{17}W_{2}$ for plane oxygen.}
         \label{fig:temp-dependence}
\end{figure}

Fig.\ \ref{fig:temp-dependence} shows, among others, the temperature
dependence of the intensity enhancement, $E$, of the $(-3/2, -1/2)$
transition due to the saturation of the $(-1/2, 1/2)$ transition.
With these values, Eq.(2) allows us to calculate $^{17}(W_{2}/W)$.  In
doing so, we have set $W_{1} = W_{2}$, which is not only reasonable,
if one assumes more or less isotropic quadrupolar fluctuations, but
also brings our result in accord with $^{17}W_{\rm eff}/{^{89}W}$ data
as we will see below.  The $^{17}(W_{2}/W)$ results are plotted
in Fig.\ \ref{fig:temp-dependence} by using an ordinate axis in such a
way that the points coincide with the bullets.  In oder to estimate
$^{17}W_{2}$ itself, we have re-measured the temperature dependence of
the nuclear magnetization recovery (obtained by standard NMR) and then
fitted the data by the theoretical expression for combined magnetic
and quadrupolar relaxation \cite{suter98} using $W_{1} = W_{2}$ and
$W$ as parameters and our values $W_{2}/W = W_{1}/W$.  The results of
this fit are presented in the inset of Fig.\
\ref{fig:temp-dependence}.

We like to stress that $^{17}W$ is different from the value of
$^{17}W_{\rm eff}$ one gets if the fit procedure uses the theoretical
expression for {\em pure magnetic} recovery ($W_{1} = W_{2} = 0$).
Lacking experimental hints, this fit procedure has been applied in all
previous HTSC oxygen relaxation studies, $i.e.$ quadrupolar oxygen
relaxation, if any, has been considered to be negligible.  Now, having
$^{17}W$ that properly represents the magnetic part of the relaxation,
we can evaluate the ratio $^{17}W/^{89}W$ and compare it in Fig.\
\ref{fig:temp-dependence} with the ratio $^{17}W_{\rm eff}/^{89}W$
from Ref.\ \cite{suter97}.  Obviously, the $^{17}W/^{89}W$ temperature
dependence becomes much weaker than the $^{17}W_{\rm eff}/^{89}W$ one;
it actually approaches the theoretical prediction [solid curve in
Fig.\ \ref{fig:temp-dependence}] of the MMP model
\cite{millis90zha96}.  This agreement is satisfying and it is based on
the assumption $^{17}W_{1} \approx {^{17}W_{2}}$ we made above.  The
$^{17}W_{2}$ results have relatively large errors as a consequence of
the large $W_{2}/W$ errors.  While the data points suggest, above
100~K, an increase of $^{17}W_{2}$ with falling temperature, also a
temperature independent $^{17}W_{2}$ is compatible with the errors.
Be that as it may, the essential result is the presence of
$^{17}W_{2}$ rather than its temperature dependence in the normal
state.

What is the origin of the quadrupolar relaxation?  A simple
phononic mechanism can be ruled out for two reasons.  First, the
temperature dependence of the quadrupolar contribution is
incompatible with a power law $T^\alpha$ with $\alpha = 2 \ldots
7$ \cite{abragam61,kranendonk54} expected for relaxation due to
phonons. Second, a dramatic softening of a phonon mode, that could
in principle lead to the observed temperature dependence, is not
taking place. Though, there is a weak softening of phononic modes
in $\mathrm{YBa_{2}Cu_{4}O_{8}}$, as observed by Raman scattering
experiments \cite{litvinchuck92}, the effect, besides occuring at
a too low temperature, is much too small to account for the
observed strong increase of the quadrupolar contribution in the
oxygen relaxation.  We can also exclude quadrupolar relaxation due
to any defect motion.  Such a motion would be clearly observed in
the $^{137}$Ba relaxation since the $^{137}$Ba quadrupole moment
is much larger than the $^{17}$O one. Yet, we did not detect such
an effect \cite{lombardi}.

A clue to the origin of the quadrupolar relaxation is provided by the
fact that below about 200~K, if the temperature decreases towards
$T_{c}$, $^{17}(W_{2}/W)$ significantly grows and then diminishes in
the superconducting state.  The increase of $^{17}(W_{2}/W)$ becomes
pronounced around a temperature $T^{\dagger}$ where we recently
detected \cite{suter97} an electronic crossover associated with
anomalies in various NMR/NQR properties.  Whether these two phenomena
are related, however, remains an open question.  We also suggested
\cite{suter97} that the discrepancy between the experimental
temperature dependence of ${^{17}W}_{\rm eff} / {^{89}W}$ and the MMP
model prediction, which does not include quadrupolar relaxation, could
be easily explained by assuming an additional quadrupolar relaxation
channel for plane oxygen.  The data of Fig.\ \ref{fig:temp-dependence}
support this idea.

Further information on the nature of $^{17}W_{2}$ is provided by the
pronounced temperature dependence of the anisotropy of the effective
relaxation rate $^{17}W_{\rm eff}$ which is not expected within the
SSF model \cite{barriquand91Horvatic93,suter97,martindale98}.
Martindale {\em et al.}\cite{martindale98} tried to explain this
dependence by assuming two spin degrees of freedom.  In contrast, the
uniform spin susceptibility monitored by the different plane nuclei
scales perfectly thus strongly supporting the view of a {\em single}
spin degree of freedom.  Therefore, we believe that, below about
200~K, the temperature dependence of the $^{17}W_{\rm eff}$ anisotropy
is caused by the additional quadrupolar relaxation channel.  If so,
this enables us to determine the anisotropy of $W_2$.

The magnitude of the $W_2$ rate itself yields an important information
on the nature of the quasiparticles.  Below 200~K, $W_2$ is much too
large to originate from simple quadrupolar relaxation due to
electron-like quasiparticles \cite{obata63obata64}.  Their fluctuation
spectral density is smeared out up to frequencies that are extremely
high as compared to the nuclear Larmor frequency; therefore, the
charge fluctuation amplitude at Larmor frequency, and consequently
$W_2$, is very small.  The substantial quadrupolar relaxation we
observe indicates that it has to arise either from strongly {\it
correlated} quasiparticles as proposed by different theoretical models
\cite{emery94ranninger96}, or from very strong electron-phonon
interactions that could lead to heavy polaronic-like quasiparticles
\cite{alexandrov94alexandrov95}.  In both cases, the slown-down
quasiparticle dynamics could produce enough low-frequency spectral
density necessary to account for the observed quadrupole relaxation.

One nevertheless would expect this relaxation to be equaly operative
at oxygen and copper sites and, due to the larger copper nuclear
quadrupole moment, even more effective at the later sites.  This is
not what one observes.  Whereas there is a substantial quadrupolar
contribution to the oxygen relaxation below 200~K, as shown in this
work, no such contribution has been detected in the copper relaxation.
Since the squared ratio of the quadrupole moments of $^{63}$Cu and
$^{17}$O is about 60, one expects that such a contribution would be
easily detected.  We have measured very accurately (by nuclear
quadrupole resonance) the ratio $R$ of the $^{65}$Cu and $^{63}$Cu
relaxation rates at the plane copper site in our sample at 100~K. Our
result, $R = 1.1497(35)$, is in the small error limits equal to the
squared ratio of the copper gyromagnetic ratios.  The high precision
of $R$ sets the limit for any possible quadrupolar contribution to the
Cu relaxation to less than 1 \% which is far from what one is
expecting.

Finally, let us return to the remark, made at the beginning of this
Letter, that antiferromagnetic fluctuations are very well filtered out
at the plane oxygen sites.  This is true if the fluctuations are
commensurate with the crystal structure.  However, recent inelastic
neutron scattering experiments revealed that there are {\em
incommensurate} antiferromagnetic correlations developing in
underdoped $\mathrm{YBa_2Cu_3O_{7-\delta}}$\cite{mook98}.  Most likely
such correlations are present also in $\mathrm{YBa_{2}Cu_{4}O_{8}}$,
so one expects much less efficient filtering of fluctuations at the
oxygen site and a more copper-like oxygen magnetic relaxation in
contrast to what is observed experimentally The discrepancy in
$\mathrm{YBa_2Cu_4O_8}$ is even more pronounced, since, so far, the
effective rate $^{17}W_{\rm eff}$ was used to analyze the data and
thus the magnetic relaxation rate $W$ got overvalued.  The
reconciliation of these NMR findings with those gained by neutron
scattering experiments poses another challenge to the theory.

In conclusion, we have shown that the spin-lattice relaxation of plane
oxygen in $\mathrm{YBa_2Cu_4O_8}$ below approximately 200~K is not
driven only by magnetic but also by quadrupolar fluctuations ($i.e.$
low- frequency charge fluctuations).  In the superconducting state,
this newly established quadrupolar relaxation diminishes faster than
the magnetic one indicating that the underlying relaxation process is
strongly influenced by the superconducting transition.  There are two
degrees of freedom involved in the low-energy excitations of the
electronic system, one of them is the single-spin degree, implying
that the single-spin fluid model is partially correct, whereas the
other one is the {\em charge} degree of freedom with predominantly
oxygen character, since it is not observed at the copper sites.

The partial support of this work by the Swiss National Science
Foundation is gratefully acknowledged.

\end{document}